\documentclass[12pt, letterpaper]{article}

\usepackage{amsmath,amssymb}
\usepackage{cite}
\usepackage{fancyhdr}
\usepackage[top=1in, bottom=1in, left=1in, right=1in]{geometry}
\usepackage{graphicx}
\usepackage{hyperref}

\numberwithin{equation}{section}
\date{}

\begin{document}
\title{{\rm\footnotesize \qquad \qquad \qquad \qquad \qquad \ \qquad \qquad \qquad \ \ \ \ \ \                      RUNHETC-2025-13
}\vskip.5in    The Hydrodynamic Approach to Quantum Gravity\\  Submitted to 2025 Gravitation Research Foundation Essay Contest March 3, 2025}
\author{Tom Banks (corresponding author)\\
NHETC and Department of Physics \\
Rutgers University, Piscataway, NJ 08854-8019\\
E-mail: \href{mailto:tibanks@ucsc.edu}{tibanks@ucsc.edu}
\\
\\
}

\maketitle
\thispagestyle{fancy} 

\begin{abstract} Several papers from the mid to late 1990s suggest that Einstein's equations should be thought of as the hydrodynamic equations of a special class of quantum systems.  A classical solution defines subsystems by dividing space-time up into {\it causal diamonds} and Einstein's equations are the hydrodynamics of a system that assigns density matrices to each diamond with the property $ \langle K_{\diamond} \rangle = \langle (K_{\diamond} - \langle K_{\diamond} \rangle)^2 \rangle = \frac{A_{\diamond}}{4G_N} $. These define the {\it empty diamond state}, the analog of the quantum field theory vacuum, in the background geometry.  The assignment of density matrices to each diamond enables one to define the analog of half sided modular flow along geodesics in the background manifold, as a unitary embedding of the Hilbert space of a given diamond into the next one in a nesting with Planck scale time steps.  We conjecture that this can be enhanced to a full set of compatible unitary evolutions on a Hilbert bundle over the space of time-like geodesics, using a {\it Quantum Principle of Relativity} defined in the text.  The compatibility of this formalism with the experimental success of quantum field theory (QFT) is discussed, as well as the theoretical limits in which QFT emerges.  This is a slightly expanded version of an essay that won Honorable Mention in the Gravitation Research Essay contest for 2025. 
\normalsize \noindent  \end{abstract}


\vspace{1cm}

\vfill\eject
\section{Hydrodynamics and Quantum Mechanics}

 The Navier-Stokes equation is a universal equation, which follows from a few general assumptions about coarse graining and symmetries in the long wave-length limit.  In classical physics it can be derived fairly rigorously from the BBGKY hierarchy and the Chapman-Enskog procedure for long wavelength approximation of the Boltzmann equation.  In quantum physics there is a long history of attempts to derive it from the Schrodinger equation, which one can find summarized in\cite{tbal}.  The result is a classical Markov equation for diagonal matrix elements of the density operator in eigenstates of the locally conserved densities, averaged over ''large" subsystems.  If the changes of these densities with time are small, the solution of this equation can be written, as first realized by M. Kac, in terms of an Euclidean path integral with a Lagrangian involving a small number of time derivatives.  
 
 In general, the variables in the hydrodynamic path integral have nothing to do with the microscopic variables of the system.  However, experience from condensed matter physics teaches us that low lying excitations of a system with an approximately degenerate ground state are best described as quantized excitations of the linearized hydrodynamic equations\footnote{Landau's Fermi Liquid theory can be viewed as a fermionic extension of this idea.  The bosonic hydrodynamic variables obey non-local equations of motion, which are simplified by the underlying fermion fields.}.  For a certain range of energies, perturbative solution of this effective field theory captures the full spectrum of the model and its transport properties.  For higher entropy states, the hydrodynamic variables give only a coarse grained description of time averaged, long wavelength physics, over time scales that are much longer than the microscopic scales in the theory, but much shorter than recurrence times (if the Hamiltonian is time independent).  A sketch of how a similar scenario could apply to Euclidean gravitational path integrals can be found in\cite{tbwormhole}.  Essentially all of the successes of these methods in calculating time dependent entropies and spectral form factors\cite{successes} can be understood in this language.  The fact that hydrodynamic path integrals are always computing time averages, rather than quantizing a model exactly, explains the puzzles with non-factorization of amplitudes that occur when non-trivial topologies are included in the path integral.
 
 The important distinction between these two uses of path integrals for hydrodynamics is not appreciated in most of the quantum gravity literature.  {\it Euclidean path integrals for hydrodynamics are a way of expressing, given a fairly mild set of assumptions, the result of classical statistical fluctuations of the coarse grained hydrodynamic fields in high entropy states, where they are not a proper description of the underlying quantum mechanics.  They should NOT be thought of as encoding non-perturbative corrections to the real time path integral description of the quantized hydrodynamic variables, which ARE the real effective low energy field theory of many important real world systems near their ground state.}  Coarse grained hydrodynamic variables are time averages and their correlation functions do not obey factorization theorems obeyed by the underlying quantum operators, to which they are only approximations.  
 
 \section{Gravity as Hydrodynamics}
 
 The {\it Covariant Entropy Principle} 
 \begin{equation} {\rm Tr} ( e^{- K_{\diamond}} K_{\diamond} ) = \frac{A_{\diamond}}{4G_N} , \end{equation} can be viewed as underlying Jacobson's\cite{ted95} hydrodynamic derivation of Einstein's equations.  It was stated as a bound by Bousso\cite{bousso}, following the generalization of the Bekenstein-Hawking entropy law to cosmological space-times by Fischler and Susskind\cite{fs}.  $A_{\diamond}$ refers to the maximal $d - 2$ volume in the null foliation of the boundary of a causal diamond, and is called the area of the diamond's holographic screen.  
 
 Jacobson's derivation uses only the Raychaudhuri equation for the expansion of a pencil of geodesics in the vicinity of a point on the holographic screen, covariant conservation of the stress tensor, and Unruh's observation that an accelerated detector in locally flat space experiences an acceleration dependent temperature.  It captures all of Einstein's equations {\it except the cosmological constant}, verifying, as confirmed by the AdS/CFT correspondence, that the c.c. is not the energy density of some particular state of the system.  
 
 The work of Carlip\cite{carlip} and Solodukhin\cite{solo} in 1998, expanded on Jacobson's work, showing that the equation for small fluctuations of the conformal factor of the transverse geometry (the geometry of the $d-2$ dimensional holographic screen) was the hydrodynamic equation for the stress tensor of a $1 + 1$ dimensional conformal field theory (CFT).  Postulating that the theory on the holographic screen was indeed such a CFT, they rederived the area law from Cardy's formula\cite{cardy}.  C-S made their observation only for black hole horizons.  It was generalized to the holoscreen of generic diamonds in\cite{BZ}, where it was also observed that it implied the general fluctuation equation
 \begin{equation}  \langle K_{\diamond} \rangle = \langle (K_{\diamond} - \langle K_{\diamond} \rangle)^2 \rangle . \end{equation}  If the central charge is large, this equation is valid if $K_{\diamond}$ is the $L_0$ generator of a cut-off $ 1 + 1$ dimensional CFT, as long as the UV cutoff is just above the range where Cardy's formula for the spectral density is valid.  The C-S ansatz implies that the central charge of the CFT is proportional to $\frac{A_{\diamond}}{4G_N} \gg 1$, in which case the spectral integral is dominated by a saddle point\cite{BZ}.  This fluctuation formula has been verified by other means for causal diamonds in a large number of different cases\cite{VZ2}\cite{deBoer}\cite{tbpdreplica}.  For large black holes in anti-de Sitter space, $AdS_d$ it is replaced by the thermal result of CFT in $d - 1$ dimensions.  The discrepancy can be explained in terms of the tensor network approximation to the AdS/CFT correspondence\cite{BZ}.  The Carlip-Solodukhin result is valid in single nodes of the network, which control the UV limit of the CFT on a sphere.  This explains why it is obtained for RT diamonds, which are UV dominated.   The entropy of large black holes is dominated by compressible hydrodynamic modes of the network, which have the fluctuation formula of a higher dimensional CFT.
 
 The tensor network approximation is also the easiest way to understand the difference between the hydrodynamic description of perturbations of black hole horizons for negative and non-negative cosmological constant\cite{MTW}\cite{liurangamani}.  As is well known, for $\Lambda \geq 0$ the hydrodynamic equations are incompressible, while for large black holes in AdS space they are compressible.  This is a consequence of the ballistic {\it vs.} fast scrambling\cite{lshpss} propagation of information for large black holes, when compared to  small black holes, non-negative c.c. or Ryu-Takayanagi (RT) diamonds.  These can all be understood in terms of the lattice structure of the tensor network.  Fast scrambling dynamics takes place within nodes of the tensor network, while large black holes are spread over many nodes and their entropy is dominated by long wavelength modes of the network.  As noted above, the entropy of infinite RT diamonds is dominated by extreme UV modes of the boundary field theory and so is sensitive to only a single node of the network.  It's the single node dynamics to which the arguments of C-S apply, as well as arguments about fast scrambling.  
 
 \section{From Density Matrices to Time Evolution}
 
 In quantum field theory (QFT) the local operator algebras of causal diamonds are Murray von Neumann Type $III_1$ and do not have density matrices, but one can always define positive definite {\it modular operators} $\Delta_{\diamond}$ for each diamond, in terms of the vacuum state.  The {\it modular flow} generated by conjugation by  $\Delta_{\diamond}^{-it}$ maps the local operator algebra onto itself.  For non-conformal field theories, it does not act locally.  Nonetheless, if we take two nested diamonds with the same past tip and future tips differing by a small amount $\delta t$ then $\Delta_{<}^{-1} \Delta_{>}$ defines a time evolution, {\it half sided modular flow}, which maps the local wedge algebra between the two diamonds into itself.  Thus, we can define a time dependent evolution inside a nested set of diamonds (Fig. 1), which is causal, in terms of the modular operators.
 
 The analogs of the modular operators in our quantum gravity formalism are the diamond density matrices $e^{- K_{\diamond}}$, and the analog of half sided modular flow is a sequence of unitary embeddings of nested diamonds along a single geodesic into diamonds one Planck step further along in proper time. This suggests that the proper framework for quantum gravity, given its hydrodynamic description in terms of a classical solution of Einstein's equations, is to describe the quantum fluctuations in terms of a {\it Hilbert bundle} over the space of time-like geodesics on the classical background.  For space-times that are asymptotically symmetric spaces in the future, with no quantum mechanically eternal black holes, the Hilbert space over each geodesic is identical, so the whole bundle is a {\it non-isometric encoding} of the quantum information in the space-time, into that single space.  In order to promote the unitary embeddings at finite times along each geodesic into a unitary operator on the full Hilbert space, we need a connection on the bundle.  Intuitively, the information that is not yet causally connected to a given geodesic, is described along a different geodesic.  The Quantum Principle of Relativity (QPR) is easy to state, but hard to implement: {\it If two diamonds have a geometric overlap, the largest diamond in that overlap should be mapped to tensor factors in both diamond Hilbert spaces.  The full unitary dynamics along each geodesic produces density matrices for those two factors, and those are required to have the same entanglement spectra.}  We conjecture that this completely determines the unitaries along all geodesics, and probably puts restrictions on which solutions of Einstein's equations correspond to valid models of quantum gravity.  At the moment we only have a very hand waving understanding of the implications of the QPR.  
 
 Fast scrambling is key to a more detailed understanding of the $ 1 + 1$ dimensional CFT that the C-S conjecture postulates on diamond boundaries.  Long before we understood the significance of the C-S papers, Fischler and the present author had postulated that Jacobson's principle be modeled by fermionic variables\cite{tbwfman}\cite{bfm}, which, in the case of the maximal entropy $p = \rho$ flat FRW cosmology, were arranged as a cutoff CFT with central charge proportional to the horizon area.  Fiol's fermionic matrix model of the Schwarzschild de Sitter entropy formula\cite{bfm} was another key ingredient in the choice of fermionic variables.  The geometric meaning of these variables, we finally understood, could be found in the work of Alain Connes\cite{connes}: the Riemannian geometry of the holographic screen of a causal diamond in a solution of Einstein's equations is completely encoded in the Dirac operator on the screen.  Solutions of a fluctuating Dirac operator can be written as sums of fluctuating two dimensional fermion fields $\psi_a (t,z)$ multiplied by eigenspinors $\chi_a (\Omega)$ of the background Dirac operator on the holographic screen.  The fields $\psi_a (t,z)$ live on an interval $z \in [0,\pi]$, with a UV cutoff on their dimensionless momentum.  The cutoff is chosen such that the thermal ensemble $e^{- L_0}$ for each field is dominated by the part of the spectrum where Cardy's formula is valid.  The central charge is determined by the C-S principle. This fixes a UV cutoff on the number of allowed transverse Dirac eigenspinors, and introduces a ''fuzziness"\cite{tbjk} into the geometry of the screen.  The smallest central charge corresponds to the place where the C-S argument, which is based on the classical Einstein equations, breaks down.  In the real world this could only be determined by experiment.  It's not clear whether any of our more precise definitions of quantum gravity in higher dimensions, which only tell us how to calculate boundary correlation functions, can determine what this cutoff should be in theoretical models.  It seems likely that there will be many local models, which give rise to the same boundary correlators\footnote{This possibility is already evident in tensor network approximations to the AdS/CFT correspondence.  Tensor networks are a sequence of lattice approximations to a CFT.  Given the Wilsonian understanding of CFTs as universality classes of long distance behavior, it seems likely that there are many consistent tensor network models that will converge to the same CFT.  }. 
 
 The free fermion CFT will not be a fast scrambler of information on the holographic screen, but there is a perturbation of it which is likely to serve the purpose.  Formally, bilinears in spinors on the screen transform as the sum of all anti-symmetric tensors under the tangent space rotation group, and thus make $p$ forms when contracted with the frame field.  Integrating products of $p$ forms over the manifold one can make interactions that are invariant under volume preserving diffeomorphisms.  These do not respect distance constraints on the manifold and should be fast scramblers.
 
 A practical way of implementing this idea uses the generalization of Fiol's\cite{bfm} matrix model idea.  The count of Dirac eigenvalues on the $d - 2$ sphere below some angular momentum cutoff is exactly the number of anti-symmetric $d - 2$ tensors on a vector space of rank $N$.  This is also the asymptotics for large eigenvalue of the Dirac eigenvalue count on any smooth manifold. Thus, we propose to introduce fermion fields
 \begin{equation} \psi_{(I(1) \ldots I(d-2))} (t,z), \end{equation} where each anti-symmetrized index $I(k)$ runs between $1$ and $N$.  We form bilinear currents
 \begin{equation}  J_{\alpha\ I}^J (t,z) = \psi^{\dagger\ (I(2) \ldots I(d-2) J)} (t,z) \gamma^0 \gamma_{\alpha} \psi_{(I \ldots I(d-2))} (t,z) . \end{equation} 
 The group of unitary transformations on the fermion fields provides a fuzzy approximation to the group of volume preserving diffeomorphisms on the manifold, and traces of powers of bilinears are the analogs of integrals of $d - 2$ forms over the manifold.  If we choose an interaction of the form
 \begin{equation} {\cal L} (t, z)  = N^{- 1 + 2 (d - 3)} {\rm Tr} [ J_{\alpha}^{(1)} J^{\alpha\ (2)}] . \end{equation}  where $J_{\alpha}^{(1,2)}$ are two commuting $U(1)$ currents chosen from among the $o(N^2)$ currents above, then in the limit where the two dimensional cutoff is removed we have a conformally invariant Thirring interaction, which is a fast scrambler on the holographic screen.
 We also want to choose the currents so that the power of $N$ in front of the interaction guarantees that the time scale for evolution is of order $N$ in Planck units, which is what we expect from the point of view of a detector on the geodesic if the system evolving lives on the stretched horizon of the diamond.  The modular Hamiltonian of the diamond is of course just the dimensionless Virasoro generator of the CFT, up to an additive constant, but this factor takes into account the red shift of times and energies.  
 
 This arrangement of degrees of freedom allows us to discuss the emergence of local physics inside a diamond.  A localized excitation ''in the interior of a big diamond of size $N$" should, at some time, be contained in a smaller causal diamond.  So we associate it with a subset of degrees of freedom $\psi_{(j(1) \ldots j(d-2))} (t,z),$, where $j(k)$ runs over a smaller number of values $n \ll N$.  Bilinears $M_j^k$ with both indices in the smaller range, are of two kinds.  They can be made purely from fermions with only small indices, or they can contain NO fermions with only small indices.   A four fermion term can also have interactions of the form
 \begin{equation} \psi_{(j(1) \ldots j(d-3), I)} \psi_{(j \ldots j(d-3))}  \psi_{(I,k(1) \ldots k(d-3))} \psi_{(k(1) \ldots k(d-3),j)} ,  \end{equation} where we've left off conjugates and two dimensional Dirac matrices.  These terms provide direct interaction between the small index variables and the rest of the system because the variables $ \psi_{(I, j(1) \ldots j(d-3))} $, are coupled to everything else.  Thus, in order for the variables $\psi_{(j(1) \ldots j(d-2))} (t,z)$ to evolve independently, we must be in a constrained state in which $ \psi_{(I, j (1) \ldots j (d-3))} (t,z) | S \rangle = 0. $  Given the scaling of the Hamiltonian, this decoupling of the localized variables can persist for times of order $N$, the time of propagation through the diamond, as viewed from the geodesic.  
 
 Note that the number of constrained qubits is of order $n^{d-3} N$, large, but much less than the total number of qubits in the system.  So our constrained state is a relatively small deviation from equilibrium.  The quantity $n^{d-3}$ should be conserved for times of order $N$, up to corrections of order $1/N$.  In the limit $N \rightarrow \infty$ it becomes an asymptotically conserved energy.  
 
 The constraints we've described on the fermion fields set the off diagonal components $M_j^I$ of bilinear matrices to zero.  We can repeat the procedure for more small groups of indices $n_i$ as long as $\sum n_i^{d - 3} \ll N^{d - 3}$, so that the constrained state is a small perturbation of the large diamond equilibrium.   This gives us a picture of multiple localized excitations evolving independently inside the large diamond, for times of order $N$.  However, now we must consider how the multiple sets of constraints are distributed among all the diamonds in a nested cover of the large diamond (see Fig. 1).

  \begin{figure}[h]
\begin{center}
\includegraphics[width=01\linewidth]{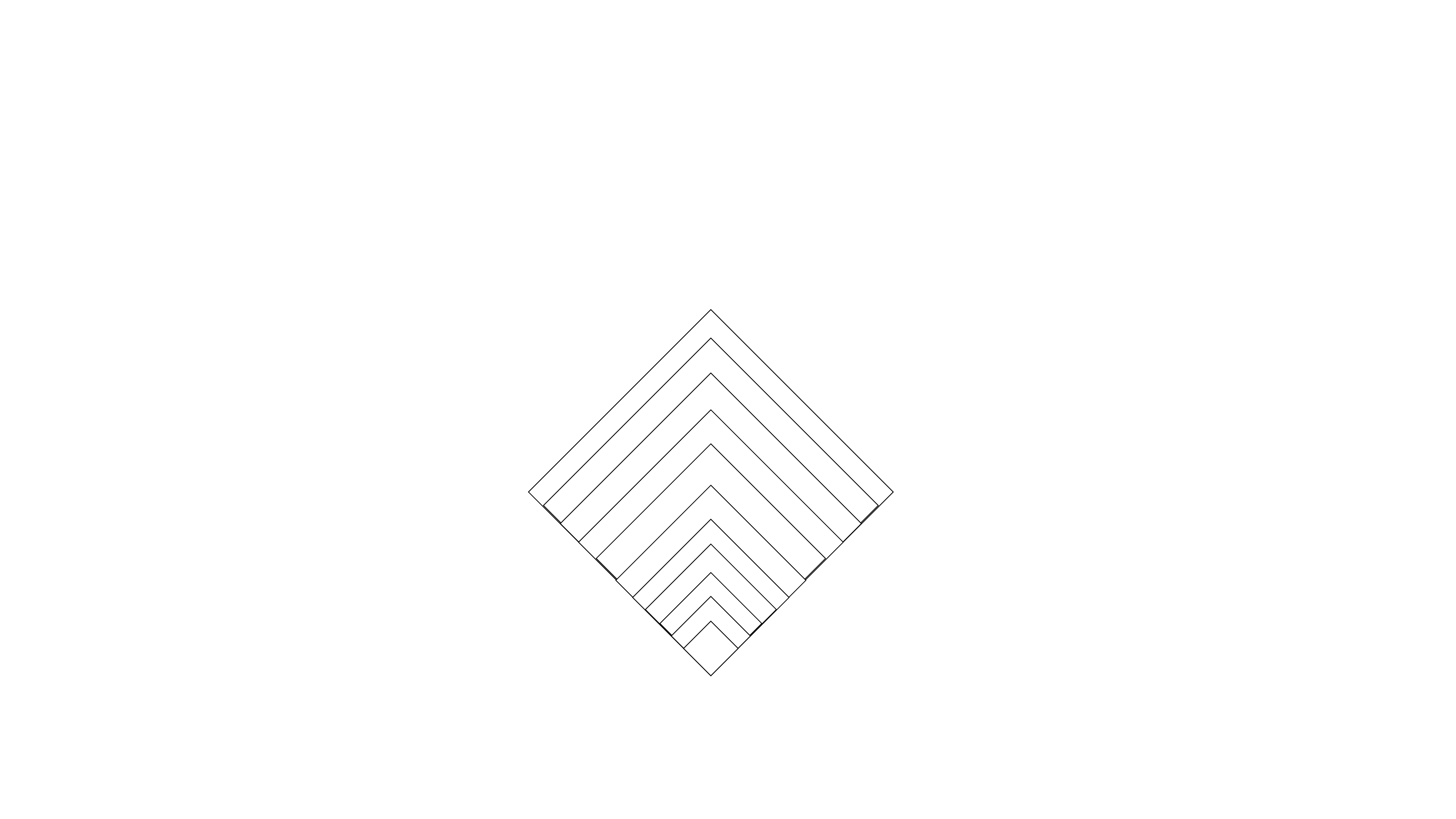}

\caption{A Future Directed Nested Cover of a Diamond. } 
\label{fig:nested}
\end{center}
\end{figure}

 If two independent subsystems of size $n_{1,2}$ both enter into an early diamond with
 $n_1^{d-3} + n_2^{d-3} \sim N_{early}^{d - 3} $ then we can no longer conclude that the final state on the future boundary of that diamond consists of two separate systems.  All we can say for certain is that it corresponds to $\sim (n_1^{d-3} + n_2^{d-3})N$ constrained fermions in the Hilbert space of the big diamond.  That might be split up into independent systems in any number of ways, but if $N_{early} \gg 1$ it is most likely to be a random state chosen from the equilibrium distribution.   
 
 We thus have a system in which initial states consisting of some number of ''independent localized excitations" can evolve into different numbers of localized excitations, or into a meta-stable equilibrium state\footnote{The equilibrium is only meta-stable because there is a finite probability that it can evolve back to a state with some number of constraints.} .  The localized excitations are characterized by an asymptotically conserved ''energy" $E_i \sim n_i^{d - 3}$, and the threshold for production of the equilibrium state of size $N$ is $E \sim N^{d-3}$.  The reader will of course notice the resemblance of these formulae to those of black hole physics.  As long as the approximately conserved energy satisfies a bound resembling the one described by the phrase "the Schwarzschild radius of the incoming energy is smaller than the size of the causal diamond", we can assign an Effective Field Theory vertex that describes the conversion of some number of localized excitations into some other number, conserving energy up to corrections of order $1/N$.
 
 We can attribute the energy deficit to energy that escapes as ``soft gravitons".  Note also that our localized excitations are much more localized than particles in QFT are allowed to be.  We attribute this to the fact that they are composite systems with of order $n^{d-3}$ q-bits.  Their energy is fixed and their direction in space-time is a collective coordinate whose quantum fluctuations decohere because they are shared by a large number of q-bits.  As a mnemonic we can call these {\it exclusive gravitational jets}.  Perhaps some connection can be made with Feynman diagram calculations involving soft graviton emission in near forward directions, but the details of this are completely unclear.
 
 Note that for $d > 4$ the bilinear matrices $M_i^j$ containing only indices in the small ranges, contain terms involving fermions with indices in the large ranges, {\it even in states where the constraints are imposed}.  Thus, there are interactions between the localized objects and the boundary degrees of freedom, even when the constraints are imposed.  We've seen however that interactions that change the number of localized particles can occur on time scales $N_{early}\ll N$, while the terms coupling localized and boundary degrees of freedom scale like $N^{3 - d} $ with $ d > 4$ and are thus negligible on time scales $N_{early} \ll N$.  In\cite{tbwfnewton} we argued that the leading order large $N$ effect of coupling to the boundary was the static Newtonian potential between pairs of localized objects. 
 
 The actual arrangement of localized objects inside a large causal diamond is determined by the QPR, applied both to a nested cover of the diamond (Fig. 1) by smaller diamonds, and by consistency of quantum information measured along different geodesics in the background space-time.   In words, constraints on the fermion variables on the past boundary of some causal diamond must match up with constraints on the variables in diamonds with which it overlaps.   For any fixed maximal proper time of all those diamonds, they are all contained in a larger diamond\footnote{Here we're actually assuming that everything is small compared to the radius of curvature defined by the c.c..}, and the constraints are all inherited from constraints on the past boundary of that diamond.  Following the constraints through the Hilbert bundle over the space of geodesics, one obtains a picture of the trajectories of localized objects through space-time (Fig. 2).
 
 \begin{figure}[h]
\begin{center}
\includegraphics[width=01\linewidth]{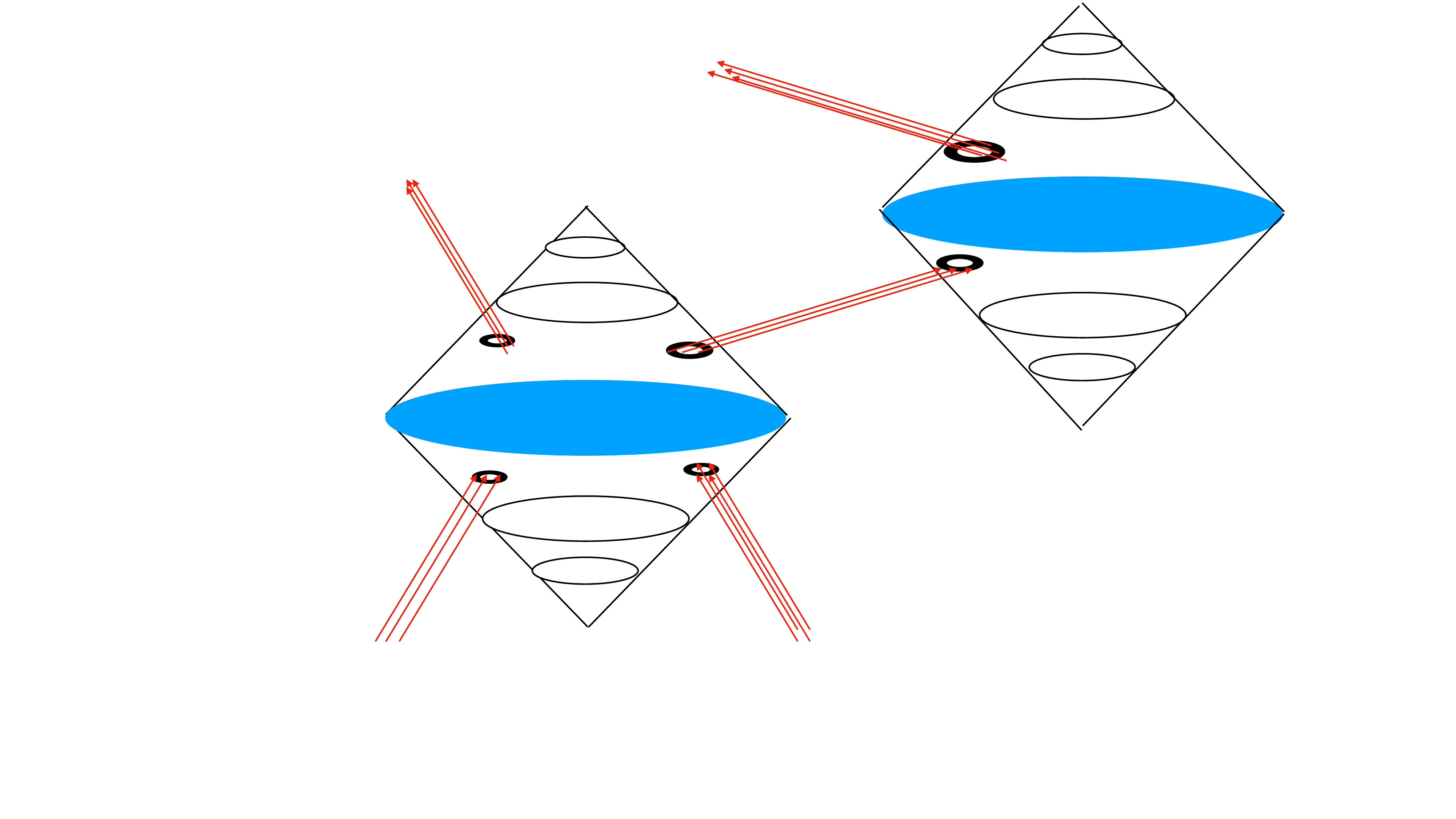}

\caption{Decomposition of amplitudes in HST models into time ordered Feynman-like diagrams, describing jets of particles propagating between causal diamonds in the background space-time. } 
\label{fig:feynman}
\end{center}
\end{figure}

 The reader will have noted how much this formulation of quantum gravity depends on the choice of background solution of Einstein's equations.  The philosophy is that the background provides a hydrodynamic approximation to the particular quantum system under study, delineating its subsystems and their density matrices in the ''empty diamond state".  Remarkably, due to the work of\cite{carlip} and\cite{solo}, combined with the principles of\cite{hst}, this leads to a guess at the underlying quantum dynamics.  It's probable that there are further constraints on the dynamics coming from enforcement of the QPR.  The biggest open question in this approach is how one constructs a mechanism like tensor calculus to guarantee that a model satisfies the QPR.  Hints from perturbative string theory suggest that the resolution of this problem might be quite subtle.  It should be emphasized that all of our experience with bona fide models of quantum gravity are in accord with the idea that it is not background independent.  Our successes consist of a collection of independent unitary S matrices or complete QFTs, each representing excitations of a particular hydrodynamic background.   Sometimes the excitations have such a large entropy that they are best treated by finding modified solutions of the hydrodynamic equations (black holes), but always the maximal entropy state is left undisturbed.  There is no evidence for, and much evidence against, the hypothesis of a background independent formulation of quantum gravity.   In this context, gravitational path integrals should be viewed in the way hydrodynamic path integrals are in CMT.  They either capture time averaged hydrodynamic fluctuations, insensitive to quantum phases, or they describe the perturbative quantum fluctuations of low energy ''phonons".   
 
 Two questions raised by this formalism are the extent to which it is consistent with the success of QFT in reproducing experimental data, and the seemingly classical nature of the trajectories of localized objects.  Both have to do with the way in which gravity upsets our preconceptions about the validity of QFT.  It has long been known that most of the states of a cutoff QFT inside a causal diamond have a gravitational back reaction which produces a black hole larger than the diamond.  Restricting attention to states that do not form such black holes, one finds that one cannot even account for the area's worth of entropy allowed by the covariant entropy bound.  One can view the ''firewall paradox" as another manifestation of the inadequacy of QFT to describe the entropy on causal diamond boundaries.   In an important paper, Cohen, Kaplan and Nelson\cite{CKN} showed that leaving out all the states that create super-diamond black holes did not affect even the most precise agreement between QFT and experiment.  Experiments are all done along a near geodesic time-like trajectory and probe only states in the vicinity of that trajectory.  If one attempts to do the same experiment simultaneously along many nearby trajectories, in order to probe a large number of states in a causal diamond, a black hole will form and the QFT description will break down along with the experiments.
 The models we have described behave like this, at least in a qualitative fashion.  
 
 The classical nature of particle trajectories has to do with the fact that in quantum gravity every particle should be thought of as an ''exclusive Sterman-Weinberg jet": a superposition of states containing the particle and any number of soft gravitons, with total momentum within some cone in momentum space.  It's clear in $4$ dimensions that the resulting state is not a normalizable state in Fock space.  The situation in higher dimensions, outside of perturbation theory is still unresolved.  What are clearly well defined in all dimensions are inclusive cross sections based on classical trajectories for the center of momentum of a hard particle and an arbitrary number of soft gravitons.  These are trajectories for high entropy states and should decohere.   The classical trajectories of localized objects in our models have the same character.

\vskip.3in
\begin{center}
{\bf Acknowledgments }
\end{center}
 The work of T.B. was supported by the Department of Energy under grant DE-SC0010008. Rutgers Project 833012.




\end{document}